\documentstyle[aps,pra,twocolumn]{revtex}
\begin{document}
\title{Reducing the communication complexity with quantum entanglement}
\author{Peng Xue, Yun-Feng Huang, Yong-Sheng Zhang, Chuan-Feng Li,\thanks{%
Email address: cfli@ustc.edu.cn} and Guang-Can Guo\thanks{%
Email address: gcguo@ustc.edu.cn}}
\address{Laboratory of Quantum Communication and Quantum Computation and \\
Department of Physics, University of Science and Technology of China,\\
Hefei 230026, P. R. China}
\maketitle

\begin{abstract}
We propose a probabilistic two-party communication complexity scenario with
a prior nonmaximally entangled state, which results in less communication
than that is required with only classical random correlations. A simple
all-optical implementation of this protocol is presented and demonstrates
our conclusion.

\end{abstract}

\baselineskip12ptQuantum mechanics provides novel features to quantum
system, extending the capabilities beyond that achievable with system based
solely on classical physics. The most prominent examples to date have been
quantum computation\cite{Shor,Grover,Ekert2,Kwiat}, quantum teleportation%
\cite{Bennett1,Pan,Boschi}, superdense coding\cite{Bennett3,Mattle}, and
quantum cryptography\cite{Ekert1,Bennett4,exp}, all of which have been
demonstrated in experiment.

Recently, there has been much interest in using quantum resource to reduce
the communication complexity\cite
{Yao2,Cleve1,Beals,Wig,Nielsen,Nisan,Dam,Klauck,Lo1}. The communication
complexity of a function $f:\left\{ 0,1\right\} ^n\times \left\{ 0,1\right\}
^n\rightarrow \left\{ 0,1\right\} $ is defined as the minimum amount of
communication necessary between two parties, conventionally referred to as
Alice and Bob, in order for both parties to acquire the value of $f$. Cleve
and Buhrman introduced the first example of the quantum communication
complexity scenario\cite{Cleve1}. In their model, Alice and Bob have an
initial supply of particles in entangled quantum state, such as EPR pairs.
They have shown that although entanglement by itself cannot be used to
transmit a classical message, it can reduce the communication complexity\cite
{Cleve2}.

In this Letter, an example of a two-party probabilistic communication
complexity scenario is presented in the entanglement model\cite{Cleve2}
which is also realized in an optical system.

Suppose Alice and Bob receive $x$ and $y$, respectively, where $x$, $y\in
U=\left\{ 0,1\right\} ^2$, and $x$, $y$ may be represented in binary
notation as $x_1x_0,$ $y_1y_0$. The common goal is for each party to learn
the value of the Boolean function 
\begin{equation}
f\left( x,y\right) =x_1\oplus y_1\oplus \left( x_0\wedge y_0\right) 
\eqnum{1}
\end{equation}
after two bits of classical communications with as high probability as
possible. If and only if the values determined by Alice and by Bob are both
correct, an execution is considered successful.

In the entanglement model, Alice and Bob initially share an entanglement of
two qubits, 
\begin{equation}
\left| AB\right\rangle =\alpha \left| 00\right\rangle +\beta \left|
11\right\rangle \text{,}  \eqnum{2}
\end{equation}
where $\alpha $, $\beta $ are promised that $\alpha ^2+\beta ^2=1$, and $%
\left| \alpha \right| >\left| \beta \right| $ (assume that $\alpha $, $\beta 
$ are real).

The idea is based on applying CHSH theorem\cite{CHSH} to enable Alice and
Bob to obtain bits $a$ and $b$ such that $a\oplus b=x_0\wedge y_0$ is
satisfied with certain probability 
\begin{equation}
\Pr \left[ a\oplus b=x_0\wedge y_0\right] =P\left( \alpha ,\beta \right) 
\text{.}  \eqnum{3}
\end{equation}
Bits $a$ and $b$ are achieved by the following operations. Suppose $R\left(
\chi \right) $ is the rotation by angle $\chi $ which is represented in the
standard basis as 
\begin{equation}
R(\chi )=\left( 
\begin{array}{cc}
\cos \chi & -\sin \chi \\ 
\sin \chi & \cos \chi
\end{array}
\right) \text{,}  \eqnum{4}
\end{equation}
if $x_0=0$, Alice applies rotation $R\left( \phi _1\right) $ on qubit $A$,
i.e.{\it ,} her part of the entangled state , else she applies $R\left( \phi
_2\right) $ on it, and then measures $A$ in the standard basis to yield bit $%
a$. Similarly, due to the symmetry of entangled states, if $y_0=0$, Bob
applies $R\left( \phi _1\right) $ on the qubit $B$, else he applies $R\left(
\phi _2\right) $, and then measures the qubit $B$ to yield bit $b$. Whereas,
local rotations $R\left( \chi _1\right) \otimes R\left( \chi _2\right) $
applied on the entangled state $\left| AB\right\rangle $ result in the state 
\begin{eqnarray}
\left| AB\right\rangle ^{\prime } &=&\left( \alpha \cos \chi _2\cos \chi
_1+\beta \sin \chi _2\sin \chi _1\right) \left| 00\right\rangle  \eqnum{5} \\
&&+\left( \alpha \sin \chi _2\cos \chi _1-\beta \cos \chi _2\sin \chi
_1\right) \left| 01\right\rangle  \nonumber \\
&&+\left( \alpha \cos \chi _2\sin \chi _1-\beta \sin \chi _2\cos \chi
_1\right) \left| 10\right\rangle  \nonumber \\
&&+\left( \alpha \sin \chi _2\sin \chi _1+\beta \cos \chi _2\cos \chi
_1\right) \left| 11\right\rangle \text{.}  \nonumber
\end{eqnarray}
After these operations, Alice sends $\left( a\oplus x_1\right) $ to Bob, and
Bob sends $\left( b\oplus y_1\right) $ to Alice, then each party can
determine the value 
\begin{equation}
\left( a\oplus x_1\right) \oplus \left( b\oplus y_1\right) =x_1\oplus
y_1\oplus \left( a\oplus b\right) =x_1\oplus y_1\oplus \left( x_0\wedge
y_0\right)  \eqnum{6}
\end{equation}
with probability $P\left( \alpha ,\beta \right) $.

The process of the communication is shown in Table I.

\begin{quote}
Table I: The input of $x_0y_0$, the corresponding local rotations, the
component of $\left| AB\right\rangle ^{\prime }$ for successful
communication and the result of Boolean function. 
\[
\begin{tabular}{|c|c|c|c|}
\hline
$x_0y_0$ & $R\left( \chi _1\right) \otimes R\left( \chi _2\right) $ & $%
\left| AB\right\rangle ^{\prime }$ & $x_0\wedge y_0$, $a\oplus b$ \\ \hline
$00$ & $R\left( \phi _1\right) \otimes R\left( \phi _1\right) $ & $\left|
00\right\rangle $ or $\left| 11\right\rangle $ & $0$ \\ \hline
$01$ & $R\left( \phi _1\right) \otimes R\left( \phi _2\right) $ & $\left|
00\right\rangle $ or $\left| 11\right\rangle $ & $0$ \\ \hline
$10$ & $R\left( \phi _2\right) \otimes R\left( \phi _1\right) $ & $\left|
00\right\rangle $ or $\left| 11\right\rangle $ & $0$ \\ \hline
$11$ & $R\left( \phi _2\right) \otimes R\left( \phi _2\right) $ & $\left|
01\right\rangle $ or $\left| 10\right\rangle $ & $1$ \\ \hline
\end{tabular}
\]
\end{quote}

According to Table I and Eq. (5), the total success probability of the
communication is 
\begin{eqnarray}
P(\alpha ,\beta ) &=&\frac 12+\frac 14\cos 2\phi _1\cos 2\phi _2+\frac 12%
\alpha \beta \sin 2\phi _1\sin 2\phi _2  \eqnum{7} \\
&&+\frac 18\left( 1-2\alpha \beta \right) \left( \sin ^22\phi _2-\sin
^22\phi _1\right) \text{.}  \nonumber
\end{eqnarray}
Then we can yield the maximum probability $P_{\max }$%
\begin{equation}
P_{_{\max }}\left( \alpha ,\beta \right) =\frac 12+\frac 14\sqrt{1+4\alpha
^2\beta ^2}\text{,}  \eqnum{8}
\end{equation}
if and only if 
\begin{equation}
\phi _1=-\frac 14\arccos \left( \frac{1+2\alpha \beta -T}{1-2\alpha \beta }%
\right) \text{,}  \eqnum{9}
\end{equation}
\begin{equation}
\phi _2=\frac 14\arccos \left( \frac{1+2\alpha \beta +T}{1-2\alpha \beta }%
\right) \text{,}  \eqnum{10}
\end{equation}
where $T=\frac{4\alpha \beta }{\sqrt{1+4\alpha ^2\beta ^2}}.$

If the two parties have previously shared an EPR pair, 
\begin{equation}
\left| \Phi ^{-}\right\rangle =\frac 1{\sqrt{2}}\left( \left|
00\right\rangle -\left| 11\right\rangle \right) \text{,}  \eqnum{11}
\end{equation}
i.e., $2\alpha \beta =-1$, the success probability of communication $P_{\max
}\left( \alpha ,\beta \right) =\frac 12+\frac{\sqrt{2}}4=\cos ^2\left( \frac %
\pi 8\right) $, and $\phi _1=-\frac \pi {16}$, $\phi _2=\frac{3\pi }{16}$,
which is just the result of Ref.\cite{Cleve2}.

This process may also be completed in another way. As we know, the
nonmaximally entangled state $\left| AB\right\rangle =\alpha \left|
00\right\rangle +\beta \left| 11\right\rangle $ (here $\beta <0$) can be
concentrated to EPR state $\left| \Phi ^{-}\right\rangle $ with probability $%
2\beta ^2$\cite{CH,Lo2}. If the concentration fails we obtain a product
state which is similar to a classical state. Whereas, with shared random
correlation instead of entanglement, the success probability cannot exceed $%
\frac 34$ (see below for detailed description). The communication is
accomplished after the concentration, therefore the total success
probability of the communication is 
\begin{equation}
P=\frac 34+\left( 2\cos ^2\frac \pi 8-\frac 32\right) \beta ^2\text{.} 
\eqnum{12}
\end{equation}
This function $P$ has a similar curve with $P_{_{\max }}$ (shown in Fig. 2),
but, obviously, the latter has higher probability, that is the protocol
presented here is superior to the protocol using entanglement concentration,
the reason lies in that some entanglement is wasted during the
unitary-reduction process of concentration.

Buhrman {\it et al.} prove that in this two-bit protocol with a shared
random correlation instead of entanglement pair, the probability is no more
than $\frac 34$ with method of ``protocol tree'' which represents any
two-bit protocol as a binary tree of depth two with non-leaf nodes labelled
A and B\cite{Cleve2}. In our case, from Eq. (8), we can verify this result
with ``limit analysis method''. If $\alpha \beta $ tends to $0$, $\left|
AB\right\rangle $ tends to classical two-bit correlation, $\left|
00\right\rangle $ or $\left| 11\right\rangle $. It is obvious that the
probability $P_{\max }\left( \alpha ,\beta \right) $ cannot exceed $\frac 34$%
.

Experimental realization of this quantum communication complexity scenario
necessitates both production and manipulation of nonmaximally entangled
states. Thus far there are only a few experimental techniques by which one
can prepare nonmaximally entangled states\cite
{Turchette,Torgerson,Digiuseppe,Kwiat1,Kwiat2}.

States with a {\it fixed} degree of entanglement, $\varepsilon \simeq \frac 4%
3$, have been deterministically generated in ion traps\cite{Turchette}, and
via {\it postselection}, nonmaximally entangled states were controllably
generated in several optical experiments\cite{Torgerson,Digiuseppe}. White 
{\it et} {\it al}. use a new kind of ultrabright source of
polarization-entangled photons to produce nearly pure entangled states
without postselection\cite{Kwiat2}, and we use the same method in this
experiment. The $\sim 1.7$-mm-diam pump beam at $351.1$ nm ($100$ mW, single
frequency) was produced by an Ar$^{\text{+}}$ laser (Sabre, COHERENT), and
directed to the two BBO crystals ($5.0\times 5.0\times 0.59$ mm), which are
aligned so that their optic axes lie in planes perpendicular to each other.
Polarization-entangled photons are produced with type I spontaneous
down-conversion process. The twin photons at $702.2$ nm are emitted into a
cone of half-opening angle $3.0^o$. Both maximally and nonmaximally
entangled states are produced simply by rotating the pump polarization. For
a polarization angle of $\chi $ with respect to the vertical, the output
state is $\left| \Psi \right\rangle =\left( \left| HH\right\rangle
+\varepsilon e^{i\Phi }\left| VV\right\rangle \right) /\sqrt{1+\varepsilon ^2%
}$, where $H$ and $V$, respectively, represent the horizontal and vertical
polarizations of two separated photons, and $\Phi $ is adjusted via the
tiltable zero-order ultraviolet quarter-wave plate (UV QWP), and the degree
of entanglement, $\varepsilon =\left| \tan \chi \right| $\cite{Kwiat2}.

In this experimental realization, an entangled pair of photons is produced
in the nonmaximally entangled state $\left| AB\right\rangle =\alpha \left|
HH\right\rangle +e^{i\Phi }\beta \left| VV\right\rangle $, where we replace
the state $\alpha \left| 00\right\rangle +\beta \left| 11\right\rangle $ in
Eq. (2) by $\alpha \left| HH\right\rangle +e^{i\Phi }\beta \left|
VV\right\rangle $, i.e., we choose horizontal and vertical polarizations as
the basis, and $\alpha =\cos \chi $, $\beta =\sin \chi $. By tilting the UV
QWP, the phase angle $\Phi $ is tuned to $0$. To show that our scenario is
universal for a complete range, a set of nonmaximally entangled states has
to be achieved by varying $\chi $ from $-45^o$ to $0^o$ (adjusting the UV
HWP). Any local rotation of the polarization (such as $R\left( \phi
_1\right) $, $R\left( \phi _2\right) $) is realized with two HWPs, whose
axes are properly oriented, in the corresponding down-conversion beam.

For each nonmaximally entangled state, there are four kinds of classical
inputs $x_0y_0=00$, $01$, $10$, and $11$, with equal probability $\frac 14$.
According to the input of $x_0y_0$, the state $\left| AB\right\rangle $ is
rotated by adjusting the HWPs in each down-conversion beam (referring to
Table I). Then with the polarizing beam splitters (PBS), the resulting state 
$\left| AB\right\rangle ^{\prime }$ can be measured to yield the bits $a$
and $b$, where $a$ $\left( b\right) =0$ or $1$, corresponding to the
horizontal or vertical polarization of each photon. The bit $a$ is detected
with detector $1$ and $2$ ($D1$ and $D2$); whereas, the bit $b$ is detected
with $D3$ and $D4$. Each detector assembly comprises an iris and a narrow
band interference filter ($702nm\pm 2nm$), to reduce background and select
(nearly) degenerate photons; a $40\times $ lens to collect the photons; and
a single-photon counter (EG\&G SPCM-AQR-16-FC), with efficiency of $\sim
70\% $ and dark count rates no more than $25$ s$^{-1}$. The detector outputs
are recorded singly, and in coincidence using a time to amplitude converter
(TAC) and a signal-channel analyzer (SCA). A coincidence window of $5$ ns
was sufficient to capture true coincidences. Compared to the typical true
coincidences of $30$ s$^{-1}$, the ``accidental'' coincidence rate is
negligible ($<0.01$ s$^{-1}$).

In this experiment, for each classical input of a nonmaximally entangled
state, we detect four coincidences of which two are corresponding to the
process of successful communications. According to Table I, if $x_0y_0=00$, $%
01$, or $10,$ it is successful communication when we detect a $D1D3$
coincidence (between detectors $1$ and $3$) or a $D2D4$ coincidence (between
detectors $2$ and $4$); whereas, if $x_0y_0=11$, it is a successful
communication when we detect a $D1D4$ or $D2D3$ coincidence. Then we can
obtain the total success probability of communication for every nonmaximally
entangled state.

With this source, we attain visibilities of better than $98\%$, when the
photons are created in the maximally entangled state. As Fig. 2 shows,
across a wide range of entanglement there is good agreement between the
experimental result of success probabilities and the theoretical predictions
of Eq. (8). The error is about $\pm 2.8\%$.

According to Table I, for the classical input $x_0y_0=11$, the output for
successful communication ,i.e., $\left| 01\right\rangle $ or $\left|
10\right\rangle $, is different from that ($\left| 00\right\rangle $ or $%
\left| 11\right\rangle $) for the classical input $x_0y_0=00$, $01$ or $10$.
If the previously shared entanglement is EPR state, the successs
probabilities for the four kinds of input are identical, i.e., equal to $%
\cos ^2\left( \frac \pi 8\right) $. However, with the entanglement decrease
to $0$, the success probability for $x_0y_0=11$ decrease gradually to $0$,
whereas, the success probabilities for $x_0y_0=00$, $01$ or $10$ increase
gradually to $1$, consequently, the overall success probability tends to $%
\frac 34$. This kind of symmetry-broken to reduce the communication
complexity is based entirely on quantum nonlocality and also been testified
in our experiment.

From Eq. (8), the success probability of communication is a function of $%
\left| \alpha \beta \right| $, and every probability corresponds to a
nonmaximally entangled pure state. The larger $\left| \alpha \beta \right| $%
, the higher probability. From this point of view the monotonicity of the
probability of communication in the present protocol may be regarded as a
kind of entanglement monotone \cite{Vidal} of a single copy of arbitrary
nonmaximally entangled pure state. We expect that quantum communication
complexity may be help to measure the entanglement in multipartitie quantum
systems.

In summary, a probabilistic two-party communication complexity scenario is
proposed and demonstrated in experiment. We showed that quantum entanglement
resulted in less communication than is required with only classical random
correlations. These results are a noteworthy contrast to actually simulate
communication among remote parties.

This work was supported by the National Natural Science Foundation of China.

{\bf Figure captions:}

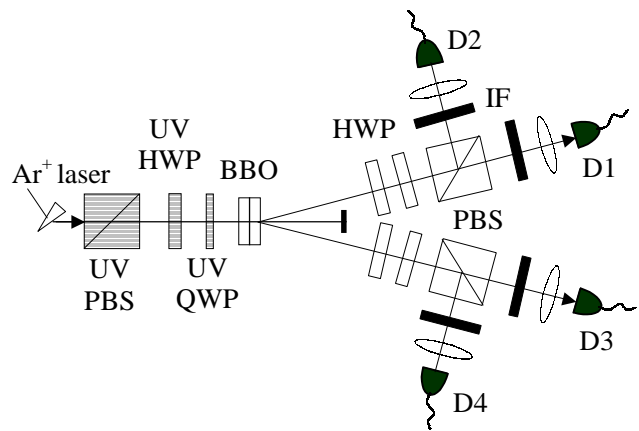
{\bf Figure 1. }Experimental setup (top view). The pump beam is polarization
filtered via a polarizing beam splitter (UV\ PBS). The polarizationof the
pump beam is set by a half-wave plate (UV HWP). The local rotation on the
entangled state is completed via two UV HWPs.

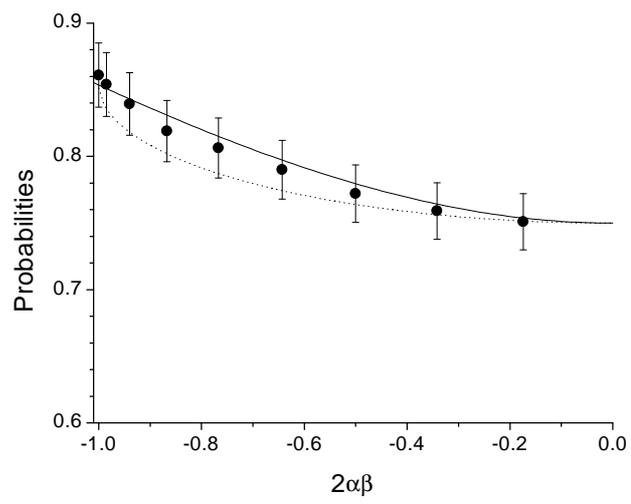
{\bf Figure 2. }The success probabilities of communication for a spectrum of
nonmaximally entangled states. {\it Points:} Experimentally determined
success probabilities, with uncertainties of $\pm 2.8\%$ (counted over 100
s); {\it Curves: }The solid line represents predicted settings for the
success probabilities in our protocol; see text for details. The dotted line
represents the success probabilities yielded by entanglement concentration.

\end{document}